\documentclass[10pt]{article}

\usepackage{amsmath}
\usepackage{amssymb}

\usepackage{graphicx}

\usepackage{cite}

\usepackage{color} 

\usepackage[labelfont=bf,labelsep=period,justification=raggedright]{caption}

\makeatletter
\renewcommand{\@biblabel}[1]{\quad#1.}
\makeatother

\date{}

\usepackage{subfigure}

\newcommand{\FRC}{FRC$^{bam}$}
\newcommand{\FRCs}{FRC$^{bam}$ }
\begin{document}

% Title must be 150 characters or less
\begin{flushleft}
{\Large
\textbf{Reevaluating Assembly Evaluations with Feature Response Curves: GAGE and Assemblathons}
}
% Insert Author names, affiliations and corresponding author email.
\smallskip\\
Francesco Vezzi$^{1,\ast}$, 
Giuseppe Narzisi$^{2}$, 
Bud Mishra$^{2,3,4}$
\bigskip\\
{\bf{1}} School of Computer Science and Communication, KTH Royal Institute of Technology, Science for Life Laboratory, 17121 Solna, Sweden 
\\
{\bf{2}} Cold Spring Harbor Laboratory, Cold Spring Harbor, United States of America
\\
{\bf{3}} Courant Institute of Mathematical Sciences, New York University, New York, United States of America
\\
{\bf{4}} NYU School of Medicine, New York University, New York, United States of America
\\
$\ast$ E-mail: Corresponding francesco.vezzi@scilifelab.se
\end{flushleft}

% Please keep the abstract between 250 and 300 words
\section*{Abstract}

In just the last  decade, a multitude of  bio-technologies and software pipelines have emerged to revolutionize genomics. To further their central goal, they aim to accelerate and improve the quality of \emph{de novo} whole-genome assembly starting from short DNA sequences/reads.
However, the performance of each of these tools is contingent on the length and quality of the sequencing data, the structure and complexity of the genome sequence, and the resolution and quality of long-range information. Furthermore, in the absence of any metric that captures the most fundamental ``{\it features}'' of a high-quality assembly, there is no obvious recipe for users to select the most desirable assembler/assembly. This situation has prompted the scientific community to rely on crowd-sourcing through international competitions, such as Assemblathons or GAGE, with the intention of identifying the {\it best} assembler(s) and their features. Some what circuitously, the only available approach to gauge \emph{de novo} assemblies and assemblers relies solely on the availability of a high-quality fully assembled reference genome sequence. Still worse, reference-guided evaluations are often both difficult to analyze, leading to conclusions that are difficult to interpret.  In this paper, we circumvent many of these issues by relying upon a tool, dubbed \FRC, which is capable of evaluating \emph{de novo} assemblies from the read-layouts even when no reference exists. We extend the FRCurve approach to cases where lay-out information may have been obscured, as is true in many deBruijn-graph-based algorithms. As a by-product, FRCurve now expands its applicability to a much wider class of assemblers -- thus, identifying higher-quality members of this group, their inter-relations  as well as sensitivity to carefully selected \emph{features}, with or without the support of a reference sequence or layout for the reads. 
The paper concludes by reevaluating several recently conducted assembly competitions and the datasets that have resulted from them.

% Please keep the Author Summary between 150 and 200 words
% Use first person. PLoS ONE authors please skip this step. 
% Author Summary not valid for PLoS ONE submissions.   
%\section*{Author Summary}
%
%\emph{Francesco Vezzi} completed his PhD in Computer Science at the Depertment of Computer Science at Udine University in Italy. Currently he is a post doctoral researcher  at KTH Royal Institute of Technology, Science for Life Laboratory in Sweden where he his main research efforts are concetrated on the design of algorithms and data structures to solve problems related to genomics with particular attention to Next Generation Sequencing technologies. He is also a member of  the Institute of Applied Genomics (IGA) in Udine.\\
%
%\noindent \emph{Giuseppe Narzisi} completed his PhD in Computer Science from the Courant Institute of NYU, where his research focused on sequence assembly algorithms, assembly validation and metrics, and accurate base-calling algorithms. He was a recipient of an IBM research fellowship. His current research focuses on algorithm design and implementations for various applications, including computational biology.\\
%
%\noindent \emph{Bud Mishra} is a Professor of Computer Science and Mathematics at the Courant Institute of NYU, where his research revolves around application of mathematics and algorithm design to  solve problems in biology, finance, economics, information, bio- and nano-technology. He is also a visiting scholar at Cold Spring Harbor Lab and an adjunct professor of NYU School of Medicine. He is a fellow of ACM, IEEE and AAAS.

\section*{Introduction}
The extraordinary advances in Next Generation Sequencing (NGS) technologies over the last ten years have triggered an exponential drop in sequencing cost, thus making it possible to perform whole-genome shotgun (WGS) sequencing of almost every organism in the biosphere.  In particular, recent WGS projects are distinctive by the way they have facilitated whole genome sequencing at a high coverage (\emph{i.e.}, higher than $50\times$), albeit, composed of relatively short sequences (\emph{i.e.}, reads). 

Despite this impressive progress, recent efforts have underlined the difficulties in trading-off read length against read coverage. It is now well recognized how the short reads have made the assembly problem significantly harder \cite{Nagarajan2009} owing to the complexity involved in resolving (\emph{i.e.} span over) long repeats. 

Nonetheless, this challenge has been confronted recently with sophisticated and novel techniques, embedded in a diverse set of tools all aiming to solve \emph{de novo} assembly problem. Such tools (\emph{i.e.},  \emph{assemblers}) are based on the simple assumption that if two reads share a sufficiently long subsequence then they are likely to belong to the same location in the genome. In order to represent and efficiently use such information for myriads of short reads, assemblers typically rely on compressed graph structures (often de-Bruijn graphs but also string-graphs). Moreover, additional heuristics are employed for error correction and read-culling. 

More than twenty different assemblers have been designed to tame the computational complexity  of assembling NGS reads, with the vast majority of them specifically targeting Illumina reads. One of the main  consequences of this proliferation in software production is the difficulty in selecting one assembler over  another, which often makes a Buridan's ass of a bioinformatics researcher: Their effort spent on selecting the best assembler (\emph{i.e.}, the largest haystack for the ass) ultimately diverts them from their ultimate objective of answering  biological questions (\emph{i.e.}, leading to a confused and starving ass). 

Adding to the confusion, every new genome presents its own sets of problems, \emph{e.g}, ploidy, heterozigosity, repetitive structures, \emph{etc.}. The available assemblers usually are able to efficiently solve only some of these problems or are specifically designed for limited datasets (\emph{e.g.}, bacterial genomes). A widely followed approach is to use multiple assemblers, run with different parameters, producing statistics that could point to the best among them. However, no clear way to select the  ``best" assembler has yet made itself obvious.
As noticed by Miller in \cite{Miller2010} all new published assemblers have been compared to the then-existing tools showing, every time, their better performances on a specific dataset and on some specific metrics.  More often than not, only  traditional metrics (\emph{i.e.}, contiguity-based metrics) are used in comparing assemblers' performances (\emph{e.g.}, number of contigs, NG50, \emph{etc.}) -- a strategy that suffers from the drawback of emphasizing only assembly size. Moreover, in \cite{Vezzi2012}, NG50 (the most ``abused'' metric) has been demonstrated to be a bad assembly quality predictor. In contrast, more reliable results can be produced, when a reference sequence is available,  since contigs could be aligned against it in order to judge the number of errors  (\emph{i.e.}, reference-based metrics). Unfortunately, currently, no effort is usually made in weighting or scoring qualitatively different types of errors, thus reducing this approach to a simple error counting without accounting for subtle differences among the different types of errors.
 
More recently, the focus has shifted from seeking just contiguity to assembly precision.  An earlier study \cite{Semple2003} showed that in the published and revised human genome \cite{Lander2001} on average 10$\%$ of assembled fragments were assigned the wrong orientation and 15$\%$ of fragments, placed in a wrong order.  Recall that this draft sequence of the Human Genome \cite{Lander2001}, which was released in 2001, had taken several large teams more than five years to finish and validate (but only at a genotypic level). With many projects left at draft level, NGS technologies have worsened this situation even further.  Alkan in \cite{Alkan2010} criticized two of the majors NGS achievements: the assembly of the Han Chinese and Yoruban individuals \cite{Li2010a} both sequenced with Illumina reads. Alkan identified 420 Mbp of missing repeat sequences from the Yoruban assembly, and estimated that in both assemblies  almost 16$\%$ of the genome was missing.

Despite these widely discussed and obvious problems, there still persists a lack of standard procedures and methods to validate and evaluate assemblies. Several projects have been initiated to explore the parameter space of the assembly problem, in particular in the context of short read sequencing \cite{Phillippy2008,Alkan2010,Narzisi2011}. 

Recently, a growing number of studies have aimed at independently evaluating different assemblers or assembly pipelines. \emph{Assemblathon 1} \cite{Earl2011} and \emph{Assemblathon 2} sought to assess assemblers' performances on common datasets encouraging a competition among researchers/users and assemblers' developers. In its earliest version, the competition was performed on a simulated dataset, leaving open  to a criticism of the effectiveness of its genome and read simulators
\cite{Vezzi2012}.  Assemblathon 1's entries were evaluated and ranked, based on a mixture of contiguity-based  and reference-based metrics. The final result is a large table (see Table 3 in \cite{Earl2011}) in which some assemblers perform well on some metrics while behaving poorly on others, thus, leaving its interpretation somewhat equivocal. 

A similar but independent study, dubbed GAGE, has been designed to critically evaluate and compare assemblers on four different large-scale NGS projects~\cite{Salzberg2011}. The presence of an already assembled reference sequence for three of the studied genomes allowed the authors to assess assembly quality. One of the main message of this study is that the same assembler can produce utterly different qualities of results on different datasets. Moreover, Salzberg and colleagues showed how assemblers' performance is affected by data quality: preprocessing used in  read correction seems fundamental  to improve assemblers' results.
The main conclusion of this study is that there is no universal ``assembly recipe" to be used for assembling new genomes.  An assembler working well on certain genomes may exhibit drastically poorer performance when used to assemble even a fairly similar genome. A fundamental criticism against GAGE  is how it selects  the ``best'' assembly for each assembler and for each dataset: GAGE's authors chose the assembly with the largest  NG50, thus building on an extremal statistic which, as mentioned earlier, also happens   to be the worst quality predictor \cite{Vezzi2012}.  As in Assemblathon 1, GAGE output is presented as  a set of tables with massive amount of---often hard-to-interpret---information. 

This state of affairs is not completely surprising, given the complexity of assembly evaluation, especially, when all errors cannot be substantially eliminated. 
For instance, even after six months since  Assemblathon 2's competition, an official ranking remains undisseminated (except for the one based on NG50).

Recently, Narzisi and Mishra in \cite{Narzisi2010} proposed a new metric, Feature Response Curve (FRCurve), capable of capturing the trade-off between contig contiguity and correctness. FRCurve is based on the principle that the assembly precision can be predicted by identifying on each contig a set of suspicious regions (\emph{i.e.}, \emph{features}): contigs are then sorted from the longest to the shortest, and for each feature threshold $\delta$ only the longest contigs whose total sum of features is less than $\delta$ are used to compute the genome coverage (\emph{i.e.}, a single point in the FRCurve).
Such technique has been extensively studied and evaluated in \cite{Vezzi2012}. Despite its power the main limitation of FRCurve is that it requires the so-called {\it read layout}, a standard output of Sanger-based assemblers, but missing in the vast majority of NGS assemblers. Such dependency restricts FRCurve analysis tools to only OLC, overlap-layout-consensus based assemblers and thus to a limited subset of NGS-based studies.  

In this paper, we present an enhanced tool, named \FRC, capable of computing FRCurve from the alignment of the reads to the assembled contigs. In particular, we show that this method is able to correctly and rigorously evaluate assemblers' performance and precision, even in the absence of a reference sequence, while using a broad set of metrics, not just those based on assembly contiguity.  
We begin by describing the set of implemented features, and then evaluate our tool on the datasets used in the three major assembly evaluations efforts:  GAGE, Assemblathon 1 and Assemblathon 2.

\section*{Materials and Methods}
Almost always, \emph{de novo} assembly is carried out using more than one library. In the Illumina scenario we typically have at least two libraries: one paired-end library (PE), and one mated-pair library (MP). The former provides paired reads in the standard orientation ($\rightarrow$ $\leftarrow $) with insert size that can vary between $150$ bp (overlapping fragments) and $1000$ bp (standard PE). The latter yields pairs of sequences in the opposite direction ($\leftarrow $  $\rightarrow$) and the insert size is much longer (usually in the range between  $3$ and $10$  Kbp).
Due to the different cost of the two protocols a typical sequencing project consists of one high coverage PE library and one low coverage MP library. The main advantage of MP reads is to improve contiguity through scaffolding and gap-filling procedures. However, the MP library is intrinsically more difficult to obtain than standard PE libraries and are usually affected by redundancy (PCR duplicates) and uneven genome representation.  
%For this reasons PE and MP reads are used to compute different features. 

After PE reads and MP reads are aligned against the assembly itself, the ordered and indexed BAM files will be the input of \FRC . \FRCs  needs at least one PE library and, if available,  one MP library.  The user needs to provide a rough estimation of the insert size and of the standard deviation for both libraries and an estimation of the genome length.  Read coverage and spanning coverage are computed directly from the BAM files.  

%\red{
Several features are computed in order to identify problems related to read coverage, mate pair happiness \cite{Phillippy2008}, and compression/expansion events (\emph{i.e.}, CE-statistics) \cite{Zimin2008}. 
%}
As a consequence of their different nature, PE reads and MP reads are used to compute two different sets of features. The former is used to compute the following features: \verb+LOW_COV_PE+, \verb+HIGH_COV_PE+, \verb+LOW_NORM_COV_PE+,  \verb+HIGH_NORM_COV_PE+, \verb+COMPR_PE+,  \verb+STRECH_PE+, \verb+HIGH_SINGLE_PE+, \verb+HIGH_SPAN_PE+, and \verb+HIGH_OUTIE_PE+.
The latter library is used to compute only a subset of the features, similar to the ones in the previous  set:  \verb+COMPR_MP+,  \verb+STRECH_MP+,  \verb+HIGH_SINGLE_MP+,  \verb+HIGH_SPAN_MP+,  and  \verb+HIGH_OUTIE_MP+.  The main difference is due to the fact that MP reads usually provide a low read coverage (\emph{i.e.} vertical) but produce a high spanning coverage (\emph{i.e.} horizontal).  Therefore MP reads are best used to compute features related to long range information  (see Table \ref{Table:FeaturesDescription} and Supplementary material for a detailed description of features).

\FRCs output consists of several files: $(a)$ the FRCurve itself (to be plotted), $(b)$ the FRCurves for each individual feature, and finally, $(c)$ a position-by-position description of the feature (in GFF format). This last file holds for each contig the identified features, together with the start and end points.

\subsection*{Datasets}
For comparative analysis of NGS assemblers, both GAGE and Assemblathon studies offer state-of-the-art datasets, which could also be re-purposed to evaluate reliability of the new \FRC.
These datasets were of particular interest to us for several reasons, falling into three categories: $(i)$ datasets consist of state-of-the-art sequences, with reads often belonging to several paired-end and mate-pairs libraries; $(ii)$  availability of already ``optimized" assemblies; $(iii)$ presence of a reference sequence for most of the sequenced organism. 

The first category allowed us to test \FRCs against state-of-the-art datasets and to take advantage of different insert-types. The second category enabled us to use assemblies that may be considered as the ``best'' achievable, since they were obtained by \emph{de novo} assembly experts (\emph{i.e.}, GAGE) or by the same assemblers' developers (\emph{i.e.}, Assemblathon).  Specifically, the availability of a reference sequence, allow us to measure assemblies' correctness, thus also demonstrating how \FRCs and the computed features are able to effectively gauge assembly accuracies and to identify suspicious regions (\emph{i.e.}, mis-assemblies).

In total, we tested \FRCs on five datasets: \emph{Staphilococcus aureus}, \emph{Rhodobacter sphaerodis}, and Human chromosome 14 from GAGE, data of simulated genomes from Assemblathon 1 competition, and \emph{Boa constrictor} (\emph{i.e.}, Snake) from Assemblathon 2 competition. All five datasets are composed of high coverage (\emph{i.e.}, all exceeding $40\times$) Illumina paired-end and mate-pair reads libraries. \emph{S. aureus} has been assembled with 7 different assemblers (see Table \ref{Table:Staph}), \emph{R. sphaerodis} and Human chromosome 14 (hereafter Hc14) have been assembled with 8 different assemblers (see Tables \ref{Table:Rhodo} and \ref{Table:HG14}). Assemblathon 1 and Assemblathon 2 comprise 59 and 12 entries respectively. The large number of Assemblathon 1 entries is simply a consequence of the rule to permit multiple submissions: we decided to download only the best entry from each team, as determined by the Assemblathon 1 ranking (refer to \cite{Earl2011} for more details), for a total of 17 entries. Summarizing, we tested  \FRCs on five extremely different datasets for a total of 43 assemblies. 
% of which 4 datasets, thanks to the availability of both a reference sequence and of an official ranking, could be used to evaluate our tool.

For each  dataset we selected one paired-end library and one mate-pair library (see Supplementary material for more details). These two libraries were then aligned against the available assemblies using rNA \cite{Vezzi2012}. We aligned reads using also BWA \cite{Li2009f} without detecting  any noticeable difference (see Supplementary material). 

Using libraries with different insert sizes (\emph{i.e.}, paired-end and mate-pair reads) enabled us to identify different features types. On the one hand, paired-end reads, characterized by a short insert size (\emph{i.e.}, usually less than $600$ bp) are able to highlight local mis-assemblies and relatively small insertions/deletions events. On the other hand, mate-pairs, characterized by a larger insert size (\emph{i.e.}, usually more than $2$ Kbp) are able to highlight larger insertion/deletion events and larger mis-assemblies (\emph{e.g.}, scaffolding errors).

\section*{Results}

Figure \ref{fig:GAGE_Assemblathon1} shows FRCurves for the three GAGE genomes (\emph{S. aureus} Figure \ref{fig:FRCstaph}, \emph{R. spheroides} Figure \ref{Table:Rhodo}, and Hc14 Figure \ref{fig:FRChg14}) and for Assemblathon 1 entries (Figure \ref{fig:FRCassemblathon1}). For each of the analyzed assemblies we aligned contigs against the reference genome. %(available in all four cases). 
To accomplish this task we employed the scripts available on GAGE website \cite{Salzberg2011}.  Assembly statistics are reported  in Tables \ref{Table:Staph}, \ref{Table:Rhodo},  \ref{Table:HG14}, and \ref{Table:Assemblathon1}.

The four tables (Tables \ref{Table:Staph}, \ref{Table:Rhodo}, \ref{Table:HG14}, and \ref{Table:Assemblathon1}) report for each assembly/assembler  the number of contigs/scaffolds produced (Ctg), the NG50, the percentage of short (\emph{i.e.} less than $200$ bp) contigs, the percentage of duplicated  (Dupl) and compressed (Comp) regions in the assembly (all the percentages are computed with respect to the real genome length), the number of long (\emph{i.e.}, $>5$ bp) indels (Indels ), and the number of Misjoins (as reported by GAGE and Assemblaton 1).  Moreover, with access to \emph{dnadiff} \cite{Phillippy2008} we could  identify regions of real mis-assemblies, thus enabling us to  compute sensitivity and specificity of our features. Note that sensitivity is defined as the ratio between true positives (\emph{i.e.},  positions marked as mis-assembled by  \emph{dnadiff} and labelled by one or more features), and its sum with false negatives (\emph{i.e.},  positions  marked as mis-assembled by \emph{dnadiff} but not labelled by any feature). Specificity, instead, is the ratio between true negatives (\emph{i.e.},  positions not marked as mis-assembled by \emph{dnadiff} and not labelled by any feature) and its sum with false positives (\emph{i.e.}, positions not marked as mis-assembled by \emph{dnadiff} but labelled by one or more features).  The first measure enables  FRCurve to identify problematic areas, while the latter measure  distinguishes non -problematic from problematic regions (\emph{e.g.}, if a feature marks all position in an assembly the sensitivity will be 1, however the specificity is likely to be close to 0).

\subsection*{GAGE}
\begin{figure}[h!]%figure1
\begin{center}
\subfigure[Staphylococcus aureus: GAGE entries]{%
            \label{fig:FRCstaph}
            \includegraphics[width=0.44\textwidth]{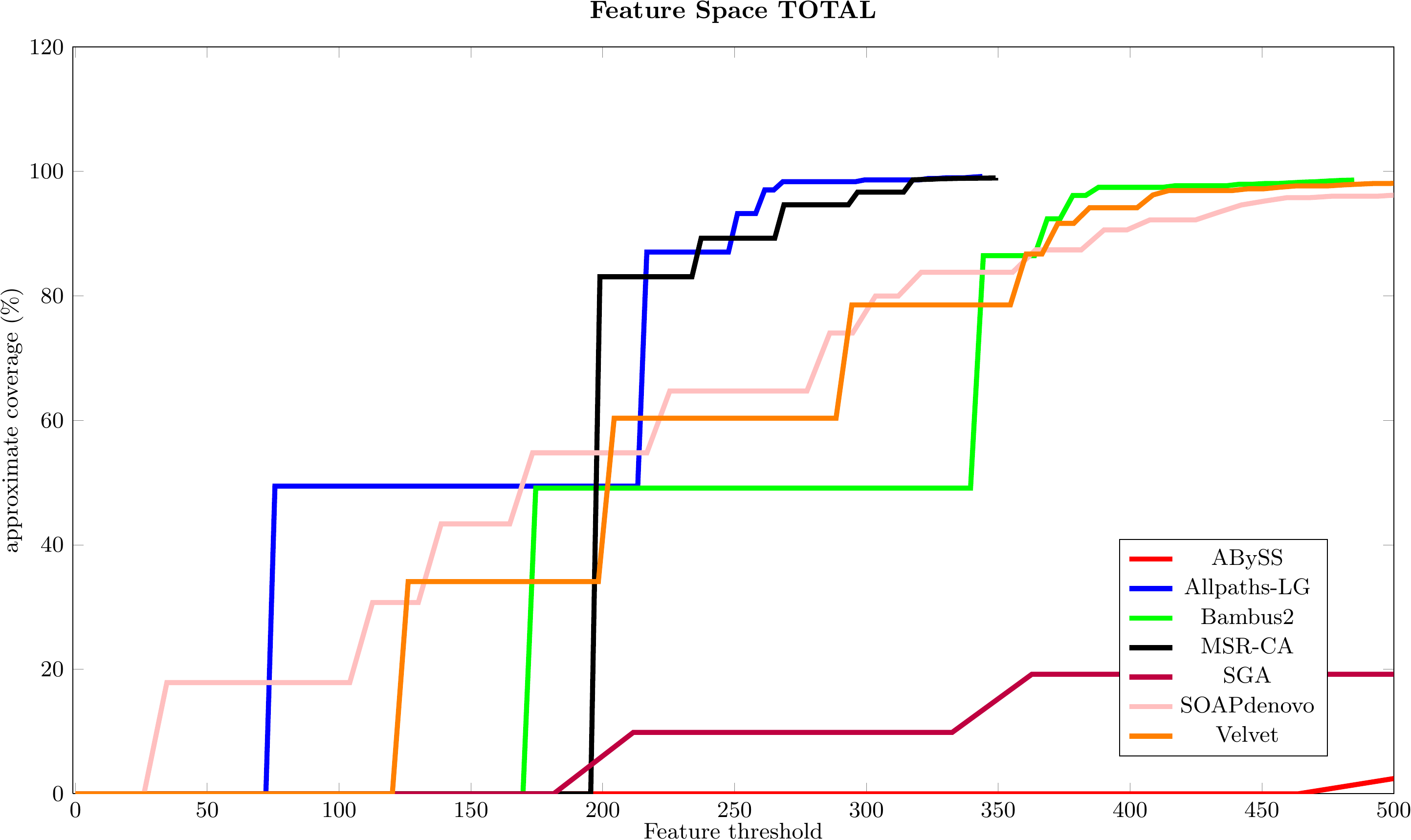}
} \subfigure[Rhodobacter: GAGE entries]{%
            \label{fig:FRCrhodo}
            \includegraphics[width=0.44\textwidth]{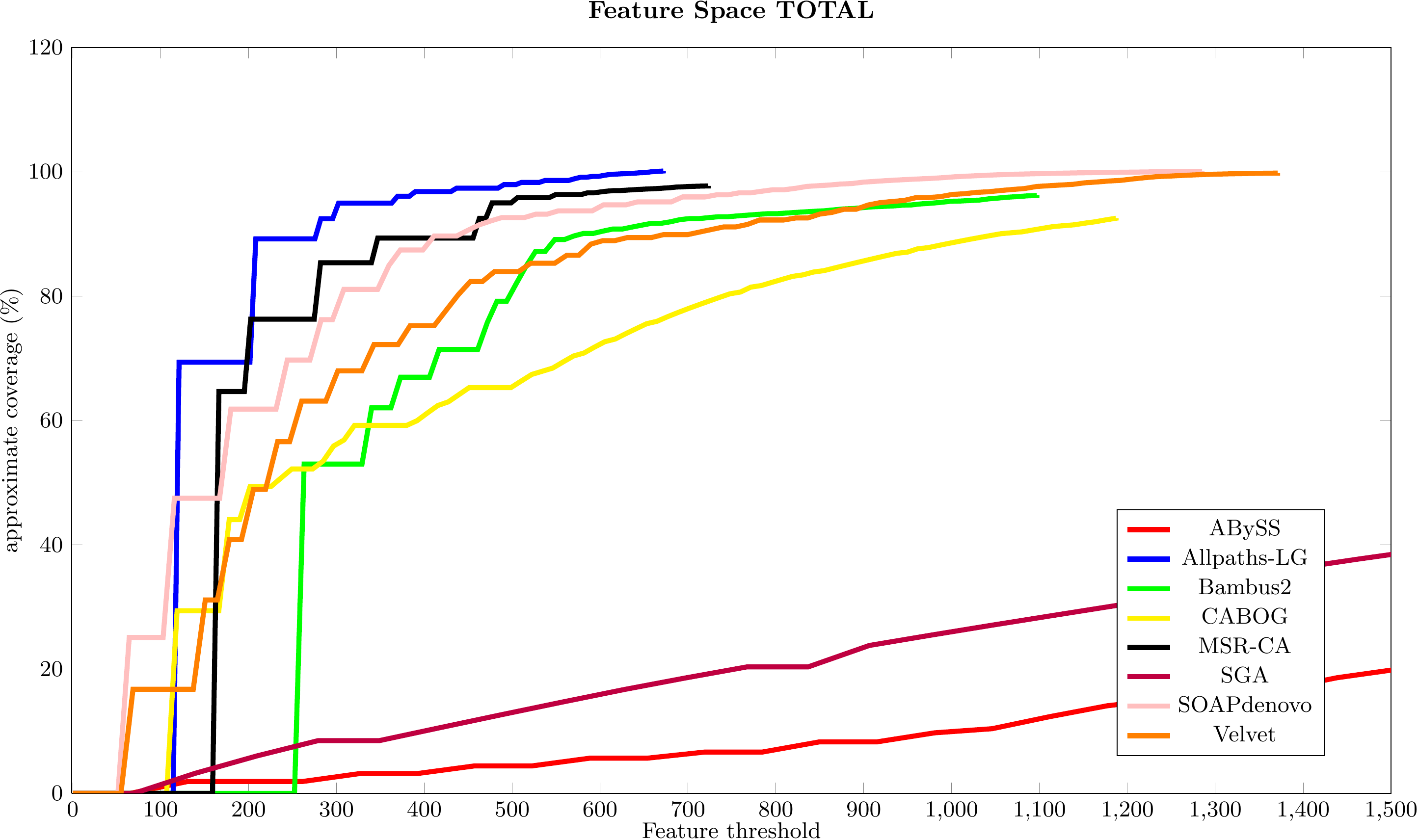}
}

\subfigure[Human chromosome 14: GAGE entries]{%
            \label{fig:FRChg14}
            \includegraphics[width=0.45\textwidth]{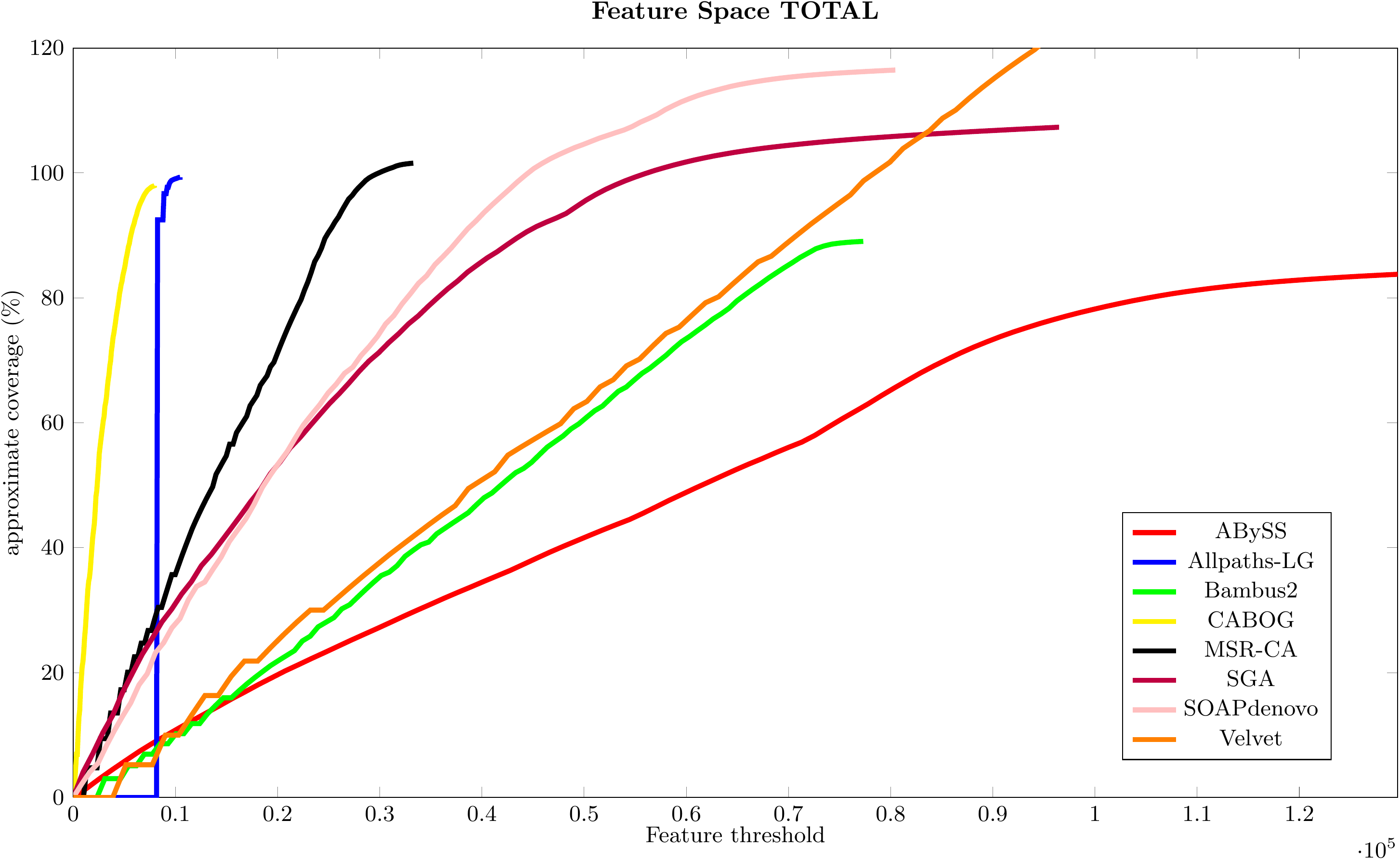}
}\subfigure[Assemblathon1 entries]{%
            \label{fig:FRCassemblathon1}
            \includegraphics[width=0.45\textwidth]{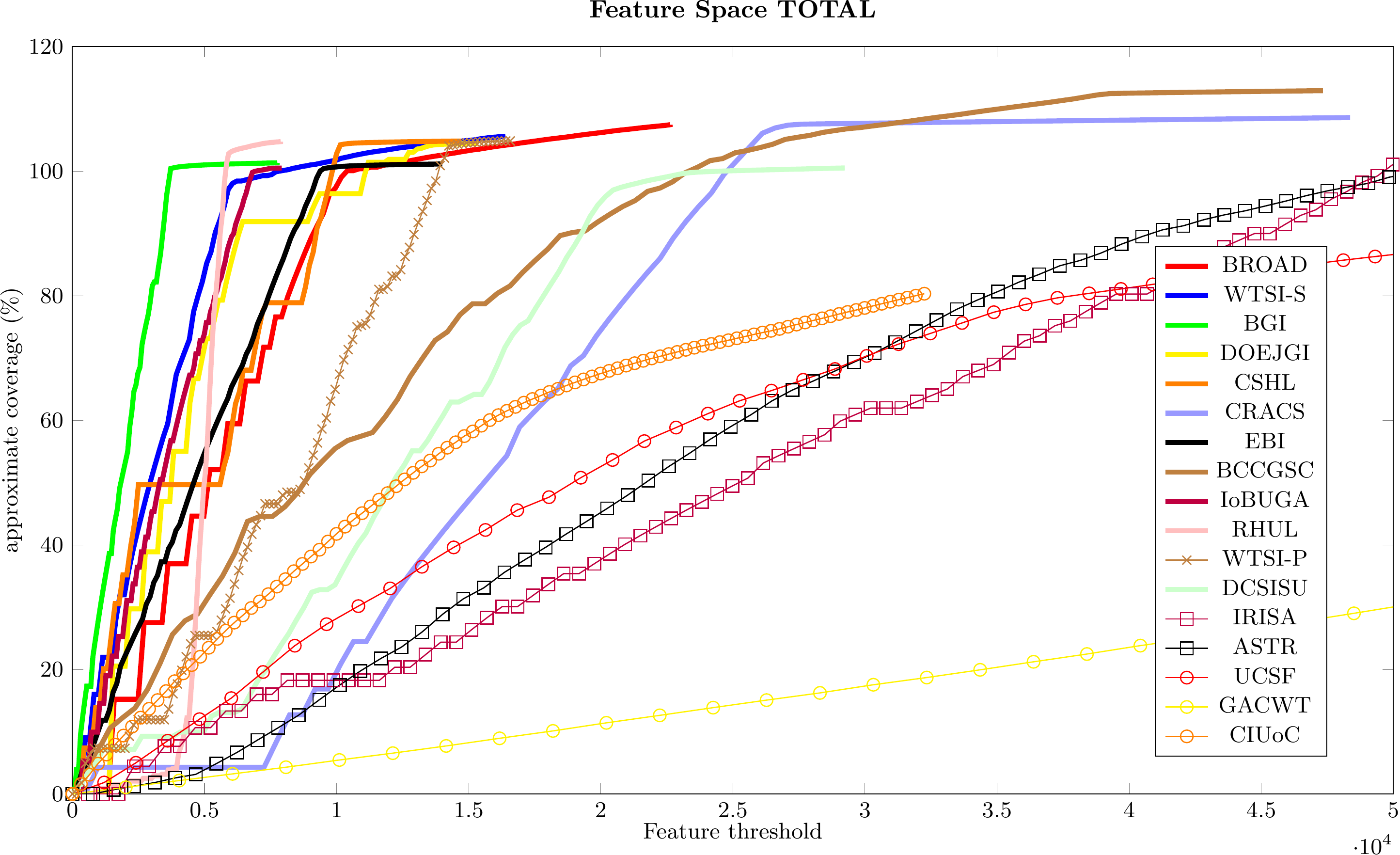}
}

\end{center}
\caption{FRCurve computed on three GAGE datasets and on Assemblathon 1 entries.}\label{fig:GAGE_Assemblathon1}
\end{figure}

Figure \ref{fig:FRCstaph} and Table \ref{Table:Staph} show the FRCurve and the reference guided validation  of \emph{S. aureus} GAGE's dataset respectively. From Figure \ref{fig:FRCstaph} MSR-CA and Allpaths-LG appear to be the best performing assemblers on such datasets (\emph{i.e.}, the sharpest curves). These two assemblers are closely followed by SOAPdenovo, Velvet, and Bambus2, while SGA and ABySS clearly show bad performance. Both sensitivity and specificity of reported features are high (last two columns of Table \ref{Table:Staph}), thus  demonstrating that \FRCs (and therefore our features) is able to correctly identify suspicious regions. Specificity is not particularly high only for ABySS and SGA. However, in these two assemblies the percentage of mis-assembled sequences identified by \emph{dnadiff} are 20\% and 8\%, respectively, suggesting a high number of problematic regions close to the real mis-assembly sequences.  

\begin{figure}[h!]%figure1
\begin{center}
\subfigure[Staphylococcus aureus vs Allpaths-LG]{%
            \label{fig:StaphDotPlotAllpaths}
            \includegraphics[width=0.44\textwidth]{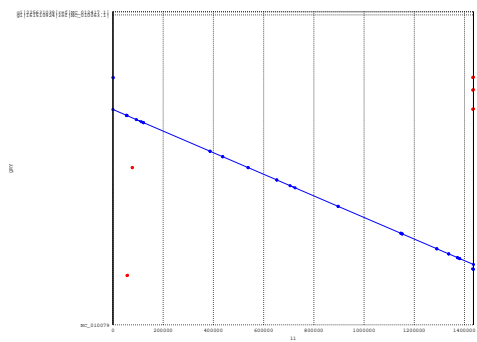}
} \subfigure[Staphylococcus aureus vs  MSR-CA]{%
            \label{fig:StaphDotPlotMSR-CA}
            \includegraphics[width=0.44\textwidth]{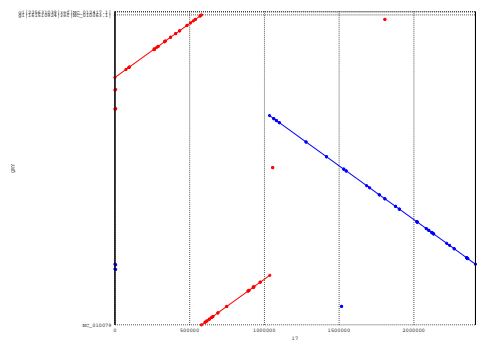}
}

\end{center}
\caption{dotPlots for Staphylococcus: MSR-CA and Allpaths-LG longest contigs have been aligned against the reference genome.}\label{fig:StaphDotPlot}
\end{figure}

Some remarks are warranted on the stepwise shape of some curves (\emph{e.g.}, MSR-CA, Allpaths-LG and Bambus2). Such a shape indicates the presence of contigs with a large number of features that interrupts a smooth growth of the curve, which is particularly discernible when the number of contigs is low. As an example, consider the longest MSR-CA contig containing almost half of the features identified in the entire assembly. The high sensitivity and specificity reported in Table  \ref{Table:Staph} show that these features represent truly problematic regions. Let us focus on Allpaths-LG and MSR-CA: in Figure \ref{fig:StaphDotPlot} we  present the alignment of the longest scaffold produced by Allpaths-LG and MSR-CA against the reference genome. From Figure \ref{fig:StaphDotPlotMSR-CA} it is clear that the stepwise shape of MSR-CA's FRCurve is a consequence of wrong choices made by the assembler. The situation is different in the Allpaths-LG case: Figure \ref{fig:StaphDotPlotAllpaths} shows a correctly reconstructed scaffold, therefore there is apparently no reason to justify the stepwise curve of Allpaths-LG. Puzzled by this anomaly, we plotted the FRCurve for each single feature (see Supplementary material). With this analysis, we discovered that Allpaths-LG has the best curve in the majority of the cases. However, there are two exceptions: \verb+STRECH_MP+ and \verb+COMPR_MP+ features, which are representative of compression or expansion events. Areas characterized by these features coincide with the circles in the dotplot (see Figure \ref{fig:StaphDotPlotAllpaths}): these areas involve small mis-joins (\emph{i.e.}, less than 50 bases) or scaffold junctions (\emph{i.e.}, sequences of Ns). A likely explanation is that such small mis-joins  are able to ``attract''  reads that are responsible for the  features. Moreover, \verb+STRECH_MP+ and \verb+COMPR MP+  features depend on CE statistics \cite{Zimin2008} and therefore on the choice of two thresholds, often estimated sub-optimally --- note that, despite the availability of a reference sequence, these thresholds were estimated without it.
MSR-CA is the assembly characterized by the largest number of areas composed of large numbers of mis-oriented mate/paired reads (\emph{i.e.}, \verb+HIGH_OUTIE_PE+ and \verb+HIGH_OUTIE_MP+),  as a consequence of the large inversions and translocations  present in the first scaffold. Other hints about MSR-CA's problems come from the FRCurve obtained from the contigs (see Supplementary material): MSR-CA´'s FRCurve is not as good as  those of Allpaths-LG, SOAPdenovo and Bambus2.

This situation demonstrates that assembly evaluation is extremely difficult. With the help of  a reference sequence it is clear that MSR-CA suffers from a large number of errors (see Table \ref{Table:Staph}). However, in its absence, many  users might have chosen  MSR-CA over others, since it seemed to be able to reconstruct almost the whole genome with a single scaffold. FRCurve, without the use of a reference, was able to raise doubts about MSR-CA (\emph{i.e.}, the only assembler with a high number of \verb+HIGH_OUTIE_PE+ and \verb+HIGH_OUTIE_MP+ features), thus suggesting a more careful manual validation on Allpaths-LG. 

According to the FRCurve analysis, SGA (together with ABySS) is one of the worst performing assemblers.  Although GAGE analysis concludes that SGA introduces relatively fewer errors,  it is also the most fragmented one, consisting of $456$ scaffolds (and $1252$ contigs).  This kind of assemblies, despite its low error-rate, tends to accumulate features related to copy number variation problems (\emph{e.g.}, \verb+LOW_COV_PE+) and features like   \verb+HIGH_SPAN_MP+ suggesting problems in the scaffolding  (\emph{i.e.}, either errors in the scaffolding or a failure in establishing contig connections).

Similar analyses can be carried out for  \emph{R. sphaeroides} and Hc14 datasets whose FRCurves  are represented in  Figures \ref{fig:FRCrhodo} and \ref{fig:FRChg14}. 

In \emph{R. sphaeroides} dataset Allpaths-LG and MSR-CA again appear to be the two best  performing assemblers, though SOAPdenovo, Velvet, and Bambus2 are not too far behind. The longest Allpaths-LG scaffold practically reconstructs the longest Rhodobacter chromosome: such scaffold contains only 100 features most of them suggesting the presence of regions affected by low paired read coverage (\emph{i.e.},  \verb+LOW_NORM_COV_PE+ and \verb+LOW_COV_PE+). Such features affect all others assemblers similarly. From FRCurve analysis one may conclude that  Allpaths-LG is  the best performing tool. The alignments of Allpaths-LG assembly against the reference further confirm this conclusion (see Supplementary  material).

Bambus2 is characterized by a long (correct) scaffold that contains almost one third  of its features. This situation is a consequence of regions composed of a large number of singleton reads (\emph{e.g.},  \verb+HIGH_SINGLE_MP+) and of areas suggesting the presence of  compression events (\emph{e.g.}, \verb+COMPR_MP+).  Similarly to the analysis of the \emph{S. aureus}, these features seem to coincide with small gaps (as the alignment of the longest Bambus2 scaffold against the reference sequence shows, see Supplementary  material). 

\begin{figure}[h!]%figure1
\begin{center}
     \includegraphics[width=0.95\textwidth]{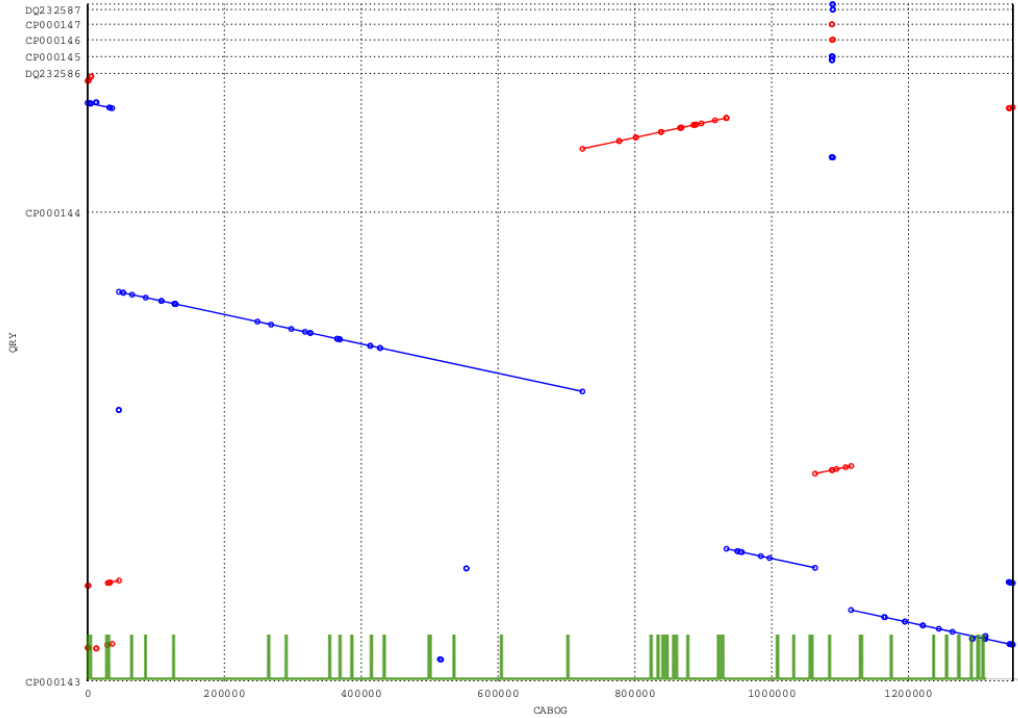}
\end{center}
\caption{Dotplot validation of the longest scaffold produced by CABOG on Rhodobacter dataset. The green line represents the Features identified by \FRC. }\label{fig:RhodobacterLongersScaf}
\end{figure}

From Figure  \ref{fig:FRCrhodo} CABOG appears not to be a very well performing assembler. Such situation is confirmed by Figure \ref{fig:RhodobacterLongersScaf} that shows the dotplot for CABOG's longest scaffolds. The green columns at the bottom of the dotplot indicate the position where one or more features have been found by \FRC. This plot shows how features are able to highlight problematic regions in the assembly, as the majority of them coincide with the mis-assemblies. 

In the Hc14 case (see Figure \ref{fig:FRChg14}) Allpaths-LG and CABOG are clearly the best two assemblers. Allpaths-LG is the only assembler able to assemble almost all the sequences in a single scaffold containing, practically, all the features. The total number of features identified on this long scaffold is lower than the total amount of features identified in the 400 longest CABOG scaffolds.  When we consider the FRCurves for each individual feature (see Supplementary material), we notice that Allpaths-LG longest contig is characterized by a large number of features suggesting coverage problems (\emph{e.g},  \verb+LOW_NORM_COV_PE+,  \verb+LOW_COV_PE+, \verb+HIGH_NORM_COV_PE+, and \verb+HIGH_COV_PE+ features) and mated/paired read orientation problems (\emph{e.g.},  \verb+HIGH_OUTIE_PE+, and  \verb+HIGH_OUTIE_MP+ features). As far as the coverage features  are concerned,  Allpaths-LG has almost always a lower number of such features than the other assemblers. Moreover, \verb+LOW_NORM_COV_PE+  feature is often the consequence of Allpaths-LG's ability to correctly resolve repeated regions (pairs are not correctly aligned as a consequence of a repeat, see Supplementary material).  
Less straightforward is the explanation for the large number of features suggesting the presence of a large number of mis-oriented pairs (in this case Allpaths-LG being one of the worst assemblers). Such features are indicative of inversions and insertions events, although the dotplot shows an almost contiguous scaffold that reconstructs the Chromosome 14 without any particular problem (see Supplementary  material).  After a closer inspection, we discovered that such long scaffold is affected by a large number of small mis-joins as suggested by the circles in the main dotplot diagonal (see Supplementary material). We tested 10 different areas subject to such mis-joins and in all cases we discovered either a  scaffold joint is too large or a scaffold joint has a short chimeric sequence in the middle, thus explaining the presence of a feature. 
The presence of these small mis-joins has been reported also in GAGE analysis: in the Hc14 dataset,  the NG50 was close to 81 Mbp while the corrected-NG50 was 20 times shorter (the corrected-NG50 is the NG50 computed after breaking contigs at mis-assembled positions identified by the reference sequence). 
The low number of compression/expansion features (\emph{i.e.}, CE statistics), as well as the low number of high-spanning and high-single reads  related features in Allpaths-LG assembly (see FRCurves plots in Supplementary material) suggests that Allpaths-LG is able to return an assembly that is highly and correctly connected. However, the relatively large number of paired-end related features suggests the presence of small local mis-assemblies. On the other hand, CABOG produced a more fragmented assembly characterized by a small number of features.  CABOG's most frequent features (\emph{i.e.},  \verb+HIGH_SPAN_PE+ and \verb+HIGH_SPAN_MP+) suggest a systematic failure during the scaffolding phase in correctly merging contigs and inferring their order.

From FRCurve analysis alone, it is much harder to decide between the top two assemblers: Allpaths-LG and CABOG, though when the reference sequence is available, it is evident that Allpaths-LG suffers less from errors than CABOG (see Table \ref{Table:HG14}). When considering only contigs (see Supplementary material) CABOG and Allpaths-LG still outperform other assemblers, as clearly proved by GAGE analysis (longest NG50).

With almost  30,000 features MSR-CA is the third ranking assembler as determined by the FRCurve analysis. MSR-CA is closely followed by SOAPdenovo and SGA. It is again difficult to fully ascertain such ranking, and even the reference guided validation in Table \ref{Table:HG14} does not lead to a clear and conclusive opinion.  The majority of SGA's features are a consequence of the highly fragmented assembly (see \verb+HIGH_SINGLE_MP+ FRCurve in Supplementary material). However the small number of errors (see Table   \ref{Table:HG14}) demonstrates that the final sequences are correct. %(may not be useful for downstream analysis). 
SOAPdenovo is slightly better than  MSR-CA as far as the number of errors is concerned, notwithstanding the fact that SOAPdenovo is more fragmented than MSR-CA. SOAPdenovo is particularly affected by the presence of mis-oriented paired reads (\emph{i.e.}, \verb+HIGH_OUTIE_PE+ feature).

In all the three GAGE datasets the sensitivity of the \FRC~is almost always higher than 90\%  (CABOG is an exception, but it must be noted that the percentage of mis-assembled sequences is less than 1.4\% of the genome length). Specificity is in general high, with the exception of assemblies characterized by high errors rates (\emph{e.g.}, more than 40\% of Velvet assembly is marked as suspicious by \emph{dnadiff} on Hc14).

\subsection*{Assemblathon 1}
Assemblathon 1 dataset differs from that of GAGE mainly in two ways: it is much larger and it is obtained solely by simulation. Figure  \ref{fig:FRCassemblathon1}  and Table  \ref{Table:Assemblathon1} summarize the analysis performed on such datasets.  It is of particular interest to compare FRCurve assembly evaluation with Assemblathon 1 paper evaluation \cite{Earl2011}. The order of the entries in Table \ref{Table:Assemblathon1}  and of the legend in Figures  \ref{fig:FRCassemblathon1} follows the Assemblathon 1 ranking.  

Despite the presence of some outliers, the FRCurve analysis is close to the ranking obtained by Earl \emph{et al.}. BGI, WTSI-S, DOEGI, and CSHL were found by the FRCurve analysis to be better performing assemblers. They, together with Broad Institute's (\emph{i.e.}, Allpaths-LG), were the five best assemblers according to Assemblathon 1 ranking. A similar analysis could determine the worst performing assemblers. CIUoC, GACWT, UCSF, ASTR, and IRISA are clearly characterized by undesirable FRCurves (CIUoC's long contigs contain few errors, even though the assembly contains only a fraction of the whole genome and small contigs contain many features). 

There are some clear differences, for example Broad's Allpaths-LG, the best assembler in Assemblathon 1 ranking is clearly among the best ones also in our FRCurve-based analysis, but has a high number of features suggesting problems with paired reads (\emph{i.e.}, \verb+LOW_NORM_COV_PE+ and  \verb+HIGH_SPAN_PE+ features). We discovered that these two features are highly correlated: in all the analyzed cases we discovered the presence of a small contig perfectly (or almost perfectly) aligning against a larger contig, probably the result of a wrong copy number estimation or of an unresolved allele splitting event. This observation is consistent with the analysis by Eearl \emph{et al.}, as, for instance, Broad's entry ranks 11\emph{th} for copy number statistics.

Another clear difference is CRACS, the 6\emph{th} ranking assembler in Assemblathon 1 evaluation, but an average performing assembler according to FRCurve analysis. The poor performance of this assembler is observed in a series of long contigs all exhibiting an extremely high coverage (\emph{i.e.}, \verb+HIGH_COV_PE+). This is clearly reflected also in the ranking given by Assemblathon 1: CRACS has clear problems in inferring copy number variation (12 \emph{th} ranking tool) and it reconstructs only $96\%$ of the genome  (14 \emph{th} ranking tool).  FRCurve analysis suggests two possible solutions: either discard contigs strongly affected by this feature, or have CRACS developers reimplement an improved copy number variation estimation.

The last two assemblers we considered are RHUL and IoBUGA. Also in this case, these assemblers have FRCurves comparable to the best assemblers, but have been ranked below the median in Assemblathon-1's evaluation. 
According to Assemblathon-1's evaluation, RHUL has an acceptable number of substitutions (5 \emph{th} ranking tool); it is able to assemble sequences in the right copy number (5 \emph{th} ranking  tool); and it is able to reconstruct (cover) the large part of the reference (4 \emph{th} ranking  tool). However, it lacks good connectivity (13 \emph{th} ranking  tool). FRCurve shows this assembler to contain most of its features in the longest scaffolds, while the short ones  contain a small number of features. Note that the longest of RHUL 's scaffolds generates a curve similar to ASTR's. 
IoBUGA offers a similar story. Assemblathon-1's ranking is difficult to interpret (15 \emph{th} ranking tool for substitutions and gene coverage but 3 \emph{rd} ranking tool for copy number variation). This situation reemphasizes that reference guided validations are extremely difficult to interpret, especially when a tool exhibits contradicting performance.  It should also be pointed out that IoBuga has the lowest sensitivity (see Table \ref{Table:Assemblathon1}).  It is clear that new features may be added in order to improve the effectiveness of \FRCs and FRCurve analysis. In this case, the availability of RNA-seq data may allow design of  new features, capable of capturing assemblers' ability to reconstruct gene expressions, splicing variants and intron-exon boundaries.

\subsection*{Assemblathon 2}
As shown earlier, the GAGE datasets were sufficient for testing the performance of \FRCs using only relatively small datasets. But with access to reference sequences, some of the limitations of the analysis became evident: only \emph{S. aureus} and \emph{R. spaeroides}  are realistic datasets, while Hc14 has been partially simulated  (reads have been aligned and extracted, see \cite{Salzberg2011} for more details). Moreover, \emph{S. aureus} and \emph{R. spaeroides} datasets are extremely small in size and, to some extent, represent  fairly easy-to-assemble genomes (\emph{i.e.}, no heterozygosity or high ploidy). With access to Assemblathon 1 data, we further tested the \FRCs against a larger dataset that was previously analyzed and ranked. The main limitation of this dataset stemmed from the use of simulated reads, which often diverged from any reasonable model of reality.

\begin{figure}[h!]%figure1
\begin{center}
\subfigure[Assemblathon 2 FRCurve: all features\label{fig:Assemblathon2_all}]{
     \includegraphics[width=0.45\textwidth]{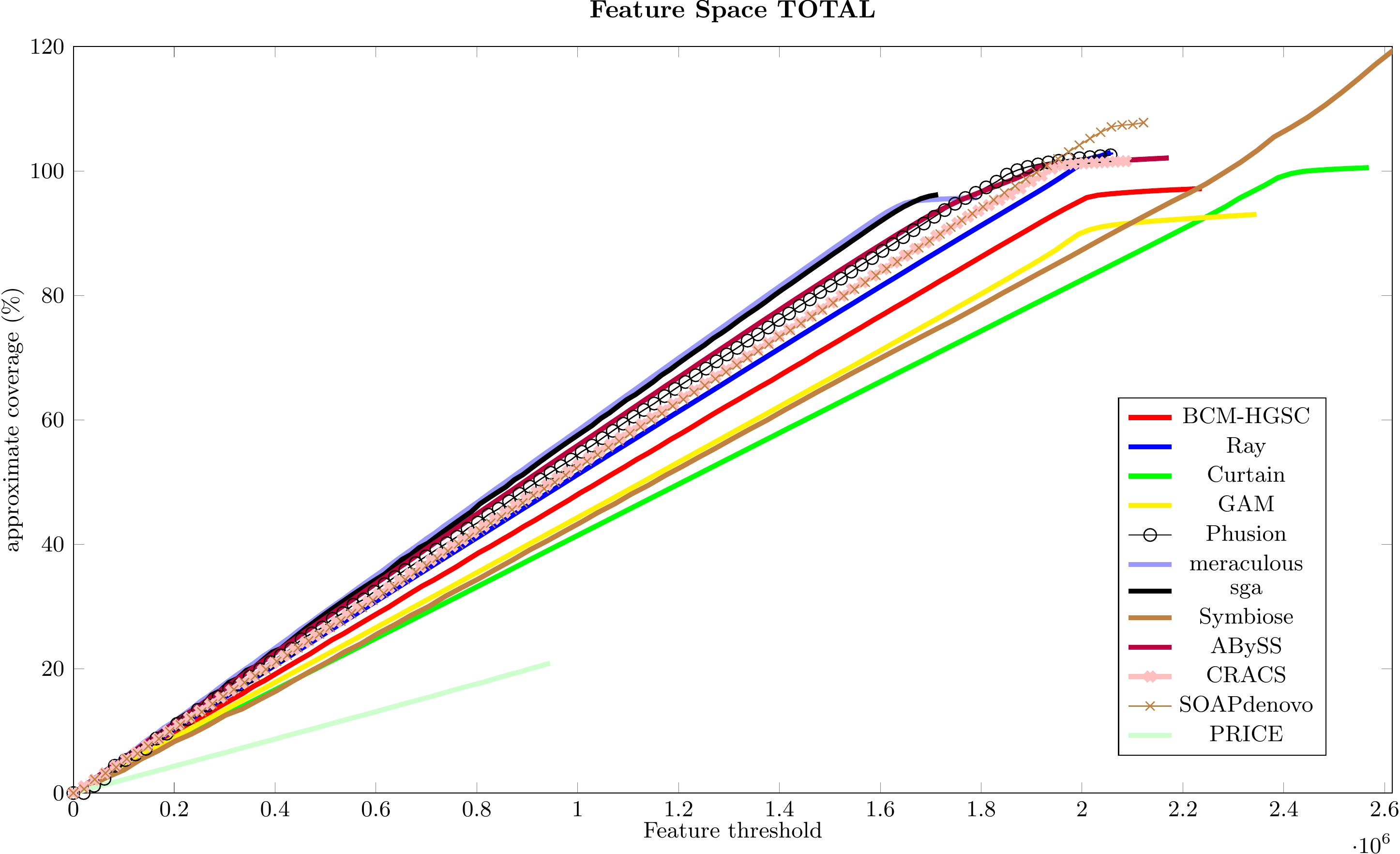}
}
\subfigure[Assemblathon 2 FRCurve: High spanning PE feature\label{fig:Assemblathon2_HIGH_SPAN}]{
     \includegraphics[width=0.45\textwidth]{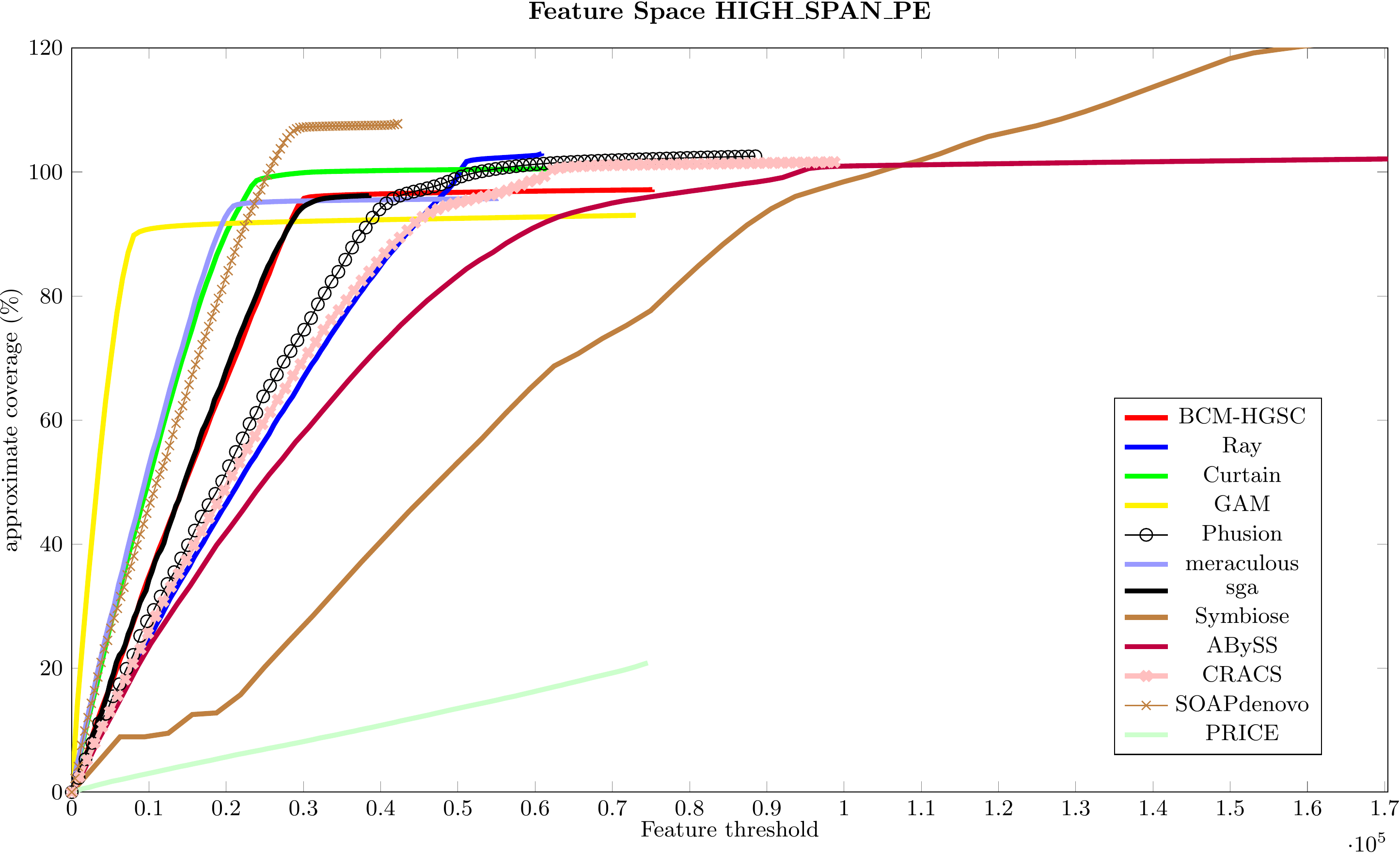}
}
\end{center}
\caption{FRCurve computed on Assemblathon 2 entries. Figure \ref{fig:Assemblathon2_all} shows FRCurves for all the features, while Figure \ref{fig:Assemblathon2_HIGH_SPAN} shows the FRCurves plotted on a single feature\label{fig:Assemblathon2}}
\end{figure}

In order to show the applicability of our method to larger sequencing projects we tested the \FRCs on all Assemblathon 2 entries for the Snake dataset (\emph{Boa constrictor}). Results are shown in Figure \ref{fig:Assemblathon2}. Surprisingly, when all features are considered all together, their FRCurves coincide closely with each other (see Figure \ref{fig:Assemblathon2_all}) suggesting that Assemblathon 2 participants, or the tools used by them, are converging to common results. We can identify two teams (assemblers) that are doing better than the others:  SGA and Meraculous. There is a dense conglomerate of similarly behaving assemblers consisting of ABySS, Phusion, SOAPdenovo, CRACS, and Ray. Other assemblers appear less promising, though, except for the sole example of PRICE, none of them show unacceptably bad performance. 
The good performance of CRACS on this dataset brings to mind how drastically differently the same assembler could behave on different datasets.

Results are different if we concentrate on one feature at a time (see Figure \ref{fig:Assemblathon2_HIGH_SPAN} and Supplementary material). As an example, by inspecting the plot for the \verb+HIGH_SPAN_PE+ feature, we observe that GAM outperforms all the other assemblers. Meraculous and SGA show good performance too, together with Curtain, Symbiose, and BCM-HGSC. \verb+HIGH_SPAN_PE+  feature indicates presence of mis-joins, as often presumably close-by pairs are found in different contigs/scaffolds. 

Particularly interesting is the FRCurve plot describing the presence of areas composed mainly of single ended reads (\emph{i.e}, \verb+HIGH_SINGLE_MP+ feature, see Supplementary material). All assemblers are strongly affected by this feature demonstrating a general failure of all tools. A likely explanation is in a systematic failure in correctly assembling heterozygous \emph{loci}, which generates holes in the assemblies, thus confounding the assemblers attempting to place both reads of mate-pairs. Note that this behavior is not present in the  \verb+HIGH_SINGLE_PE+ features. A feature like \verb+HIGH_SINGLE_MP+ is clearly not informative in this dataset and may be ignored without affecting the analysis.

\section*{Discussion}

\subsection*{Limitations of de novo assembly evaluation}
The rapidly growing set of new assemblers aims to address the need for assembly tools capable of handling  the vast amount of data produced by NGS (\emph{e.g.} Illumina) sequencers. This growth in data and tools, however, has led to another unmet need: a rigorous comparative study of these assemblers, which so far has only been carried out in a rather na\"ive way. Developers have focused more on performance (\emph{e.g.}, RAM and CPU time) and connectivity (\emph{e.g.}, contig number and NG50) rather than on correctness.

A commonly employed approach, currently being used to validate and gauge assemblies, is based on a \emph{plethora} of \emph{standard validation metrics}. We can identify four main groups: length-base statistics, reference-based statistics,  simulation-based statistics, and  long-range-information (LRI) based statistics.

Length-based statistics take into account only the size of the assembler output. These statistics comprise mean contig length, maximum contig length, and NG50.  
%NG50 represents the size $N$ such that $50\%$ of the genome is contained in contigs of size $N$ or greater\footnote{In literature, some times $NG50$ is used to indicate the largest contig such that the sum of all the contigs larger than it is at least half of the genome length, while $LG50$ indicates the size of the $N5G0$ contig. However, in our experience the definition provided in the text is the most widely used.}.
NG50, in principle, gives an idea of assemblies'  connectivity level. \emph{All} length-based statistics are not linked to assembly correctness and emphasize only length: an assembler that eagerly merges together contigs can produce assemblies characterized  by a large NG50 and by few long contigs. However, these long contigs are of no use if they contain too many misassemblies. NG50 has been shown  in \cite{Vezzi2012} to be a bad quality predictor. Nevertheless, length-based statistics are the basic, and some times the only, method used to judge assemblers performances, especially when the assembly tools are new~\cite{Zerbino2008,Reinhardt2008}.

Assembly analysis would trivialize if the genome to be assembled was already available, which would make it possible to compare assemblers using only the reference-based statistics. The strategy would be to resequence an organism with an already available fully finished whole genome reference sequence. This approach would enable comparing assemblers from the computed real number of errors. The underlying premise  is that  good performances of an assembler on one dataset should reflect behavior on  a wider range of datasets. However, studies like GAGE has shown that the same assembler can produce utterly different results on different genomes and different datasets --- thus dashing any hope of generalizing the performance of a tool on the basis of a single dataset. Moreover, reference-metrics are in general difficult to interpret or, at least, are open to several interpretations: as an example reference-based metrics have been used both to demonstrate the high quality assembly of two human individuals in \cite{Li2010a} as well as to demonstrate the opposite (their poor quality) in  \cite{Alkan2010}.

Simulation-based statistics face even more extreme hurdles: reads are simulated from a reference sequence and subsequently assembled. Vezzi \emph{et al.} showed in \cite{Vezzi2012} that simulated reads are likely to produce unrealistic contigs that cannot be used to judge assemblers' performance. Despite these shortcomings, competitions like Assemblathon 1 have continued to use a simulation-based approach.

A more reasonable way to assess assembly correctness consists in the use  of long range information.
%, better if independent from the assembly (\emph{i.e.}, not used in the assembly process). 
Second Generation Technologies are able to produce \emph{mate-pairs}, that are pairs of reads at a mean distance of $2-8$ Kbp. Mate-pairs play a crucial role in contig scaffolding, but they can be also used to gauge the assembly correctness: pairs should map on the assembly at the estimated distance and with the right orientation (depending on the sequencing technology being used). If such data is not used at assembly time it can be used as an external proof of correctness. A similar approach has already been applied with success in \cite{Phillippy2008} (\emph{i.e.}, mate-pair happiness). Other two commonly used LRI-methods are \emph{physical maps} \cite{Brody1991} and \emph{optical maps}  \cite{Anantharaman97, AnantharamanMS97}. Both rely on the relative locations of different genes and other DNA sequences of interest in the genome. Third Generation Sequencing Technologies (also known as Single Molecule Sequencing Technologies) and dilution-based sub-genomic sampling can also be used in the near future to estimate assembly correctness.
%This kind of data allows estimating the correctness of the assembly.  
The main drawback of LRI statistics is the fact that they require the production of new and often expensive data. Moreover, apart from the simple counting, it remains unclear how such information should be used to rank different assemblies that currently exist.

\subsection*{FRCurve}
The aim of this work is to present a new simple tool able to accurately evaluate assemblies and assemblers' performance even in the absence of a reference sequence. Features have been first introduced in \cite{Phillippy2008} to identify possible mis-assemblies. Narzisi and Mishra \cite{Narzisi2011} used such features to compute the so called Feature Response Curve (FRCurve). FRCurve is closely connected to the  standard receiver operating characteristic (ROC) curve: the Feature-Response curve characterizes the sensitivity (coverage) of the sequence assembler output (contigs) as a function of its discrimination threshold (number of features/errors). Given a set of features, the response (quality) of the assembler output is then analyzed as a function of the maximum number of possible errors (features) allowed in the contigs. More specifically, for a fixed feature threshold $\tau$, the contigs are sorted by size and, starting from the longest, only those contigs are tallied, if their sum of features is less than $\tau$. For this set of contigs, the corresponding approximate genome coverage is computed, leading to a single point of the Feature-Response curve. 

Vezzi \emph{et al.}~\cite{Vezzi2012} analyzed Feature space using multivariate techniques (\emph{i.e.}, PCA and ICA) in order to study features' interactions and to use these to select the most important ones. Such study, however, highlighted one of the main weak points of FRCurve: the need of a layout file, that is, a file describing the positions and orientations of  each read (and therefore, each pair). While this file had been standard with old Sanger-based assemblers, only a small fraction of NGS-based assemblers provide such information (\emph{i.e.}, Velvet, Ray, Sutta). Another relevant problem, deeply connected to the first, is the fact that features were computed by \emph{amosvalidate}. Such features are commonly available for Sanger reads, clearly characterized by widely-varying insert-size distributions and expected coverages.

Results summarized in this paper clearly show that \FRCs is able to effectively detect mis-assemblies and that it is able to rank assembler performances. The tool achieves high sensitivity and high specificity thus demonstrating that the implemented features are able to capture the large majority of the problems. Currently 9 features are computed using reads from paired-end libraries, while other 5 are computed using reads from a mate-pair library. FRCurve is computed using all of them, however the user is free to concentrate only on a subset of them (PCA can be used as shown in \cite{Vezzi2012} to study features, see Supplementary material). New forensics features can be easily added to the program in order to highlight new problematic regions: small indels can be identified using reads aligned with gaps (\emph{i.e.}, reads aligned with Smith-Waterman-like algorithm), problems in reconstructing gene space can be identified using RNA-seq reads, physical-maps  or long single-molecule-sequences can be used to compute features, highlighting scaffolders' performance.

Mapping reads back to the assembly provides only  a rough approximation of the layout generation, especially in presence of repeat-structures: in such cases, reads that belong to correctly (or incorrectly) reconstructed duplicated regions can only be mapped randomly on one of the possible occurrences, thus, jeopardizing the hope of obtaining a correct layout. \
%\textcolor{red}{
FRCurve's ability to detect mis-assemblies is clearly limited by the presence of non-uniquely aligning reads (\emph{i.e.}, reads  aligning optimally in two or more positions). Thus, as the repetitive structures in a genome increase, which complicates the assembly problem, so does the difficulty in providing valid assembly evaluation. As the read-lengths increase or mate-pairs of different lengths become feasible, not only does the assembly problem become more tractable, but also new features enable better identification of problematic regions.
%}

%\textcolor{red}{
Despite the severe limitations imposed by the strategy of approximating read layout with read alignment,
%} 
the present trend suggests that assemblers may continue to avoid producing layout files. Thus, it is believed that \FRCs and, more in general, forensics features, will need to be computed by mapping reads back to the assembled sequence. The approach to approximate the layout by mapping reads back to the assembly has several advantages: $(i)$ possibility to scale to any genome size (\FRCs is currently being used to evaluate Spruce genome assembly, which will produce a reference genome of length 20 Gbp); $(ii)$ possibility to compute new forensics features; $(iii)$ study relationships among features in a more uniform way.

%\textcolor{red}{
%In same sense, assembly validation is problem as difficult as \emph{de novo} assembly itself. 
%None of available assemblers is currently able to address all the problems related to \emph{de novo} assembly. 
Thanks to the feature-by-feature analysis, the FRCurve is often able to express and explain the current limitations of different assemblers.
%showing how different assemblers present  complementary ``expertises'' on different datasets. 
In many situations it is straight-forward to rank the assemblers simply by inspecting the FRC curves. Even when the scenario is unclear, FRCurve is still useful to highlight advantages and disadvantages of one assembler over the other (\emph{e.g.}, an assembler that presents good long range connectivity  but makes many mistakes in the small contigs, versus an assembler that has low connectivity but does not present local mis-assemblies). It is important to recall that, currently, none of the standard \emph{de novo} evaluation metrics is able to capture these situations in the absence of a reference sequence.
%}.

We believe that features-based analysis will guide efforts aimed at \emph{de novo} assembly evaluation and \emph{de novo} assembler design. Our results clearly show that FRCurve can easily separate the best  assemblies from the worst ones. By comparing feature-specific curves one can evaluate strong and weak points of each assembler and choose the system that best fits one's objective. %Assembly/assembler ranking is still difficult, however, for the first time, \FRCs has provided results that allow systematic evaluation of assemblies and assemblers even in absence of a reference sequence or a layout. 
It is hoped that, in future, assembler-developers will be guided by the features-based analysis to improve these tools --- at the core of the current genomic revolution.
% :  by analyzing FRCurve for each single feature, implementers can better understand the limits of their tools. 
%For example, Assemblathon 2 dataset showed us how a feature affects all the assemblers, thus highlighting a problematic aspect of the dataset that no existing tool was able to solve.  

\section*{Software and Data Availability} 
The sequencing data used in this study is publicly available on the GAGE website and on the Assemblathon website (details are available in Supplementary material). \FRCs source code can be downloaded from \texttt{https://github.com/vezzi/FRC\_align.git}

\section*{Supplementary Material}
All supplementary material is available at:

\texttt{http://www.nada.kth.se/$\sim$vezzi/publications/supplementary.pdf}\\
and

\texttt{http://cs.nyu.edu/mishra/PUBLICATIONS/12.supplementaryFRC.pdf}

% Do NOT remove this, even if you are not including acknowledgments
\section*{Acknowledgments}
We would like to thank  all the Spruce Assembly Project, in particular Prof. Lars Arvestad,  Bj\"{o}rn Nystedt and Nathaniel Street for their constant feedback and advice. We also wish to thank Mike Schatz of CSHL, Mihai Pop, XXX for many useful comments on an earlier draft of the paper.
Moreover, FV would like to thank Knut and Alice Wallenberg Foundation for their support. The research reported in this paper was partially supported by an NSF CDI grant.

%\section*{References}
\bibliography{FRCurve.bib}

\section*{Figure Legends}

\clearpage
\section*{Tables}

\begin{table}[h!]
\begin{tabular}{l|l}
Feature & Description\\\hline
\verb+LOW_COV_PE+                   &   \emph{low read coverage}  areas (all aligned reads).\\
 \verb+HIGH_COV_PE+                 &   \emph{high read coverage} areas (all aligned reads). \\
 \verb+LOW_NORM_COV_PE+       &   \emph{low paired-read coverage} areas (only properly aligned pairs).\\
 \verb+HIGH_NORM_COV_PE+      &   \emph{high paired-read coverage} areas (only properly aligned pairs).\\
 \verb+COMPR_PE+                      &   \emph{low CE-statistics} computed on PE-reads.\\
 \verb+STRECH_PE+                     &   \emph{high CE-statistics} computed on PE-reads.\\
 \verb+HIGH_SINGLE_PE+             &   \emph{high number of PE reads with unmapped pair}.\\
 \verb+HIGH_SPAN_PE+                &   \emph{high number of PE reads with pair mapped in a different contig/scaffold}.\\
 \verb+HIGH_OUTIE_PE+               &   \emph{high number of mis-oriented  or too distant PE reads}.\\
 \verb+COMPR_MP+                      &  \emph{low CE-statistics} computed on MP reads.\\
 \verb+STRECH_MP+                     &   \emph{high CE-statistics} computed on MP reads.\\
 \verb+HIGH_SINGLE_MP+             &  \emph{high number of MP reads with unmapped pair}.\\
 \verb+HIGH_SPAN_PE+                &   \emph{high number of MP reads with pair mapped in a different contig/scaffold}.\\
 \verb+HIGH_OUTIE_PE+               &   \emph{high number of mis-oriented  or too distant MP reads}.\\
\end{tabular}
\caption{Description of implemented features.\label{Table:FeaturesDescription}}
\end{table}

\begin{table}[h!]
\begin{small}
\begin{tabular}{lccccccccccc}
assembler 	      &  Ctg      &  NG50         & Chaff          & Dupl       & Comp        & Indels & Misjoins & Inv    & Reloc & Sens     & Spec \\
 	      &        &    (Kbp)    &  (\%)  &  (\%) &  (\%) &  &  &     &  &      &  \\\hline
ABySS                &   246     &   34             &  6.66 &  23.06 &  1.05   &  10    &    6       &  4       &     2    &  99.25  &  62.70  \\ 
Allpaths-LG       &   \bf{12} &  1,092        &  0.03 &  0.03  &  1.26    &  12    &    \bf{4} &  \bf{0} &     4    &  84.79  &  89.97  \\ 
Bambus2           &   17        &  1,084        &  0.00 &  0.01  &  1.27    &  215  &   14       &  2        &   12    & 97.14  &  83.51  \\ 
MSR-CA             &   17        &  \bf{2,412} &  0.00 &  0.80  &  0.89    &  14    &   15       &   9       &   6      & 88.12  &  92.89  \\ 
SGA                   &   546       &  208          &  0.00  &  0.02  &  1.27    & \bf{4}&   \bf{4}  &  1        &  \bf{3}& 95.48  &  63.71  \\ 
SOAPdenovo      &   99         &  3312        &  0.35  &  1.42    &  1.39  &  36    &   25      &  2        &  23     &  95.32  &  86.69  \\ 
Velvet                &   45          &  762         &  0.41  &  0.09  &  1.29    &  16    &   31      &   10     & 21       &  96.83  &  84.26  \\ 
\end{tabular}
\end{small}
\caption{\emph{Staphilococcus aureus} (GAGE) assembly evaluation and features estimation. For each assembler we report the number of contigs/scaffolds produced (Ctg), the NG50, the percentage of short (Chaff) contigs, the percentage of duplicated  (Dupl) and compressed (Comp) regions in the assembly (all the percentages are computed with respect to the real genome length), the number of long (\emph{i.e.}, $>5$ bp) indels (Indels ), the number of Misjoins, the number of inversions  (Inv),   the number of relocations (Rel), the features sensitivity (Sens), and the features specificity (Spec).\label{Table:Staph}}
\end{table}

\begin{table}[h!]
\begin{small}
\begin{tabular}{lccccccccccc}
assembler 	      &  Ctg      &  NG50         & Chaff          & Dupl       & Comp        & Indels & Misjoins & Inv    & Reloc & Sens     & Spec \\
 	      &        &    (Kbp)    &  (\%)  &  (\%) &  (\%) &  &  &     &  &      &  \\\hline
ABySS           &   1701  &  9              &  1.59  &  9.93  &  0.49 &  38    & 24   & 2         & 22      &   98.92  &  37.26  \\
Allpaths-LG  &\bf{34}  &\bf{3,192}  &  0.01  &  0.49  &  0.44 &  37    &  6     & \bf{0} & 6 &  90.73  &  93.36  \\
Bambus2      &   92      &  2,439       &  0.00  &  0.00  &  0.25 &  378  &  5     &  0       & 7        &  75.84  &  82.76  \\
CABOG         &   130    &  66            &  0.00  &  0.12  &  0.71 &  24    & 15    &  5       & 10      &  89.04  &  82.51  \\
MSR-CA       &   43      &  2,976       &  0.00  &  1.05  &  0.53  &  31   & 15     &   3     & 12      &  87.87  &  93.92  \\
SGA              &   2096  &  51           &  0.00  &  0.05  &  0.98  &\bf{4} &\bf{4} &  \bf{0}& \bf{4}  &  96.66  &  62.89  \\
SOAPdenovo &   166    &  660         &  0.44  &  1.06  &  0.53  &  431  & 11     &  1      &  10    & 92.90  &  86.62  \\
Velvet           &   178    &  353         &  0.48  &  0.29  &  0.97  &  27    & 21     &  6      &  15     &  92.04  &  83.33  \\
\end{tabular}
\end{small}
\caption{\emph{Rhodobacter sphaeroides} (GAGE) assembly evaluation and features estimation.  For each assembler we report the number of contigs/scaffolds produced (Ctg), the NG50, the percentage of short (Chaff) contigs, the percentage of duplicated  (Dupl) and compressed (Comp) regions in the assembly (all the percentages are computed with respect to the real genome length), the number of long (\emph{i.e.}, $>5$ bp) indels (Indels ), the number of Misjoins, the number of inversions  (Inv),   the number of relocations (Rel), the features sensitivity (Sens), and the features specificity (Spec).\label{Table:Rhodo}}
\end{table}

\begin{table}[h!]
\begin{small}
\begin{tabular}{lccccccccccc}
assembler 	      &  Ctg      &  NG50         & Chaff          & Dupl       & Comp        & Indels & Misjoins & Inv    & Reloc & Sens     & Spec \\
 	      &        &    (Kbp)    &  (\%)  &  (\%) &  (\%) &  &  &     &  &      &  \\\hline
ABySS           &   51301   &  2,1              &  34.78  &  0.48  &  0.44  &  762     &   \bf{22} & \bf{15} &  \bf{7} & 95.83  &  18.79  \\
Allpaths-LG  &   \bf{225}&  \bf{81,647} &  0.02    &  0.22  &  2.08  &  2575   &   146      &  44       &  102    & 68.46  &  96.79  \\
Bambus2      &   1792     &  324             &  0.00    &  0.12  &  3.33  &  5651   &   3409    & 1759     & 1650  & 86.26  &  55.04 \\
CABOG         &   479       &  393             &  0.00    &  0.16  &  1.38  &  2894   &   746       & 435      & 311    & 62.19  &  95.92   \\
MSR-CA       &   1425     &  893             &  0.01     &  1.19  &  2.15  &  3097  & 2311       & 83       & 1439    &  86.10  & 84.71 \\
SGA              &   30975   &  83               &  0.00    &  0.13  &  1.78  &\bf{681}& 150         & 90         & 60     &   92.13  &  65.38   \\
SOAPdenovo &   13501   &  455             &  3.09    &  5.68  &  3.19  &  3902   &  1529      &  537      &  992  & 90.59  &  73.10  \\
Velvet           &   3565     &  1,190          &  4.23    &  0.08  &  0.53  & 4172    &  9525      & 4023     & 5502  &    91.60  &  67.55  \\
\end{tabular}
\end{small}
\caption{Human chromosome 14 (GAGE) assembly evaluation and features estimation.  For each assembler we report the number of contigs/scaffolds produced (Ctg), the NG50, the percentage of short (Chaff) contigs, the percentage of duplicated  (Dupl) and compressed (Comp) regions in the assembly (all the percentages are computed with respect to the real genome length), the number of long (\emph{i.e.}, $>5$ bp) indels (Indels ), the number of Misjoins, the number of inversions  (Inv),   the number of relocations (Rel), the features sensitivity (Sens), and the features specificity (Spec).\label{Table:HG14}}
\end{table}

\begin{table}[h!]
\begin{small}
\begin{tabular}{lccccccccccc}
assembler 	      &  Ctg      &  NG50         & Chaff          & Dupl       & Comp        & Indels & Misjoins  & Sens     & Spec \\
 	      &        &    (Kbp)    &  (\%)  &  (\%) &  (\%) &  &  &     &  &      &  \\\hline
BROAD    &   989      &\bf{8,396}& 0.00   &  3.54    &  0.63    & 903        &  236       &  92.99  &  93.88 \\
BGI          &   1897    &  1,716     & 0.26   &  1.10    &  0.46    & 994        & 656        &  81.39  &  97.48 \\
WTSI-S    &   1380   &  2,874      &  0.00   &  0.98   &  0.98    & \bf{132} & 197        &  95.10  &  96.55   \\
DOEJGI    &   771     &  9,073      &  0.03   &  0.01   &  0.84    & 163         & \bf{181}&  94.32  &  96.80\\
CSHL       &   1842   &  3,254      &  3.05   &  0.33   &  0.66    & 3704       & 733       &  90.76  &  95.18  \\
CRACS    &   6165    &  2,712      &  0.00   &  0.56   &  0.92    &  319        &  990     &  96.59  &  83.27  \\
BCCGSC  &   3314    &  825         &  2.92   &  6.62   &  0.83    &  488        &  636     &  96.69  &  88.97 \\
EBI          &   2173    &  959         &  0.39    &  0.08  &  0.84    &   674       &  1021   &  78.66  &  94.24 \\
IoBUGA    & \bf{467} &  1,801     &  0.18   &  0.51   &  0.74    & 3596        & 1249    &  71.65  &  94.16   \\
RHUL       &   4999    &  43          & 0.00   &  0.50    &  0.84    & 336          & 1040    &  91.70  &  95.88  \\
WTSI-P    &   1448   &  502         &  0.00   &  0.34   &  0.32    &  4121       &  2389   &  93.53  &  89.94 \\
DCSISU    &   4790   &  315         &  0.00   &  0.59   &  1.22    & 1284        &  2366   &  90.14  &  79.60 \\
IRISA       &   3539   & 1,406        &  0.05   &  0.52   &  0.82    &  2518       & 350      &  95.28  &  76.90  \\
ASTR       &   6228   &  57            &  0.00   &  13.16 &  0.84   &  336          &  2265   &  91.79  &  69.97 \\
UCSF       &   14821  &  22           & 0.00   &  24.91  &  1.33    & 12131      &  5127   &  93.85  &  66.19  \\
GACWT    &   24297  &  9             & 0.00   &  24.88  &  1.03    & 2197       & 1487    &   94.10  &  49.36  \\
CIUoC      &   14993  &  6             & 0.00   &  0.09    &  1.13    & 3215       & 1889      &  77.09  &  67.29  \\

\end{tabular}
\end{small}
\caption{Assemblathon 1 assembly evaluation and features estimation.  For each assembler/team we report the number of contigs/scaffolds produced (Ctg), the NG50, the percentage of short (Chaff) contigs, the percentage of duplicated  (Dupl) and compressed (Comp) regions in the assembly (all the percentages are computed with respect to the real genome length), the number of long (\emph{i.e.}, $>5$ bp) indels (Indels ), the number of Misjoins, the features sensitivity (Sens), and the features specificity (Spec).\label{Table:Assemblathon1}}
\end{table}

\end{document}